\documentclass[twocolumn,prl]{revtex4}%

\pdfoutput=1

\usepackage{epsfig}
\usepackage{graphics}

\begin{document}
\bigskip
\hskip 4in\vbox{\baselineskip12pt \hbox{FERMILAB-PUB-13-580-A}  }
\bigskip\bigskip\bigskip

\title{ Quantum Indeterminacy of Cosmic Systems}

\author{Craig J. Hogan}
\affiliation{  University of Chicago and Fermilab}

\begin{abstract} 
It is shown  that  quantum uncertainty of motion in  systems controlled mainly by gravity generally grows with orbital timescale $H^{-1}$,   and dominates classical  motion for trajectories separated by distances  less than  $\approx  H^{-3/5}$ in Planck units. For example, the cosmological metric today becomes indeterminate at macroscopic separations, $H_0^{-3/5}\approx 60$ meters. Estimates suggest that    entangled non-localized  quantum states of  geometry and matter may significantly affect fluctuations during inflation, and connect the scale of   dark energy to that of strong interactions.
 \end{abstract}
\maketitle

\section{introduction}

Gravitational systems are  described  by classical dynamics of Newtonian mechanics and general relativity, in which trajectories are determinate paths in  a continuous dynamical manifold of space and time.  However, in reality those  systems are not entirely classical:  gravity couples to matter, which is governed by  quantum dynamics.  Systems of matter and energy are essentially  indeterminate,  and transformations of quantum states  are not generally localized in space and time\cite{zeilinger1999}. 

In many situations,  quantum behavior is confined to small scales,
 so that the classical description of space-time and gravity can be  applied to macroscopic systems. This approximation works well on scales much larger than the Planck length,  for systems dominated by  forces much stronger than gravity\cite{wilczek1999}.  However, it is shown here  that the scale of indeterminacy in  systems controlled mainly by gravity grows with gravitational timescale.  As a result, in  systems with very small mean space-time curvature and mass density, such as the universe,
 orbital motion is indeterminate  even on  macroscopic scales.  This unfamiliar behavior, due to the weakness and universality of gravity, modifies the conventional separation of quantum and classical motions. 
 It follows from just standard non-relativistic quantum mechanics and gravity.

 In general relativity,  orbits of matter trace geodesics of the metric and determine the sources of  metric curvature, so  indeterminate motion implies an indeterminate metric.   Because the relations used here assume a classical space-time and non-relativistic kinematics, they break down as the quantum-classical boundary and horizon scale are approached, so from these arguments alone we cannot  elucidate the  nature of quantum-geometrical degrees of freedom, or precisely address relativistic effects. 
  However, it is argued here that  the magnitude of the  uncertainty is large enough that observable behavior of some systems,  such as cosmic expansion driven by dark energy\cite{weinberg89,Frieman:2008sn} or early-universe inflation\cite{Lidsey:1995np,Lyth:1998xn},  may depend  on new kinds of entanglement between quantum states of matter and geometry that are not included  in standard quantum field theory.

\section{Indeterminacy in Gravitating Systems}

It is convenient to express  mass and length in  Planck units, {\it i.e.},  $m_P\equiv \sqrt{\hbar c/ G}= 1.22\times 10^{19} {\rm GeV/c^2}$, $l_P\equiv ct_P\equiv \sqrt{\hbar G/c^3}= 1.616\times 10^{-35}$m, where $G$ denotes Newton's constant, $\hbar$ denotes Planck's constant, and $c$ denotes the speed of light. In the universe today, the Hubble time is $c/H_0\approx 1.3 \times 10^{26}{\rm m}= 8\times 10^{60}$.  Because the goal is to estimate general scaling behavior, we omit numerical factors of the order of unity  that depend on details of system configuration.

\begin{figure}[b]
 \epsfysize=3.5in 
\epsfbox{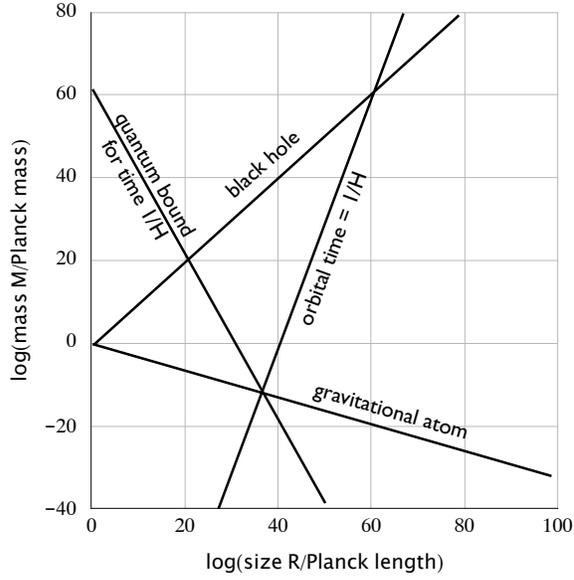} 
\caption{ \label{qc}
Relations of system mass and size  from 
Eq. (\ref{bohr}) (gravitational atom),
Eq. (\ref{systemgravity}) (orbital time), and
Eq. (\ref{quantumbound}) (quantum bound).  The  relations scale for any spacetime curvature $H^2$, but  are plotted here using  the cosmic expansion rate today, $H= H_0\approx 10^{-61}$.  Quantum  indeterminacy exceeds classical cosmic motion  in any region smaller than the intersection of these lines at $H^{-3/5}$, or about 60 meters for the current cosmic expansion. Systems above the  black hole relation, $M=R$, are unphysical. } 
\end{figure}

Consider first  two  neutral  particles of equal mass $M$ whose motion obeys the Schr\"odinger equation.  The stationary ground state of this simple quantum system is the analog of an atom, but with  gravitational binding instead of electricity.
The size $R$ of the ground state wave function is estimated by equating gravitational energy with kinetic energy in a matter wave  with momentum $\approx \hbar/R$, 
\begin{equation}
GM^2/R  \approx  \hbar^2/R^2 M.
\end{equation}
Expressed in Planck units, the size of the stationary state is given by a gravitational Bohr radius, with  mass taking the place of electric charge:
\begin{equation}\label{bohr}
R\approx M^{-3}.
\end{equation}
This relation is shown in Figure  (\ref{qc}).   It defines a minimum size for a stable self-gravitating system,  described not by classical gravitational dynamics, but by a   quantum wave function of this scale. The orbits of the particles, as well as the gravitational potential shaped by them, are indeterminate quantum objects.  

This example shows that  macroscopic quantum indeterminacy appears in gravitating systems with small mass  and large size. Indeed,  a gravitational atom made of neutron-mass particles would  be almost as large as our universe.   
Such systems were contemplated long ago by Weyl, Eddington and Dirac.

Gravitational atoms do not exist in the real world, but we now show that a similar indeterminacy  applies generally to orbits in low-density  systems of large mass, using cosmic expansion as an example.
Consider  a system of two  bodies of equal mass $M$ separated by $R$,  in a universe  with classical  expansion velocity  $v=HR$.    We ask what are the constraints on   $M$ and $R$, such that the change of position of the two bodies due to  classical expansion is greater than that due to their gravity, and also greater than quantum measurement uncertainty.  The arguments below apply  not just to cosmology, but to any gravitational system with Riemann curvature of order $H^2$.

The gravitational free-fall time  of the bodies exceeds $H^{-1}$ if they contribute less mass than an upper bound
 \begin{equation}\label{systemgravity}
M<R^3H^2.
\end{equation}
The space-time curvature  associated with the gravity of the bodies  is then less than $H^2$.  Any measurement  of system trajectories must use masses below this relation, shown in  Figure  (\ref{qc}), so that their gravity does not dominate the motion (and increase the curvature) of the cosmic system under study.  

The two bodies can also be considered as a quantum-mechanical system. In this case the separation $\hat x$ and relative velocity $\hat p/M$ are described by conjugate operators whose values are indeterminate.  Their wave functions obey  the standard Heisenberg uncertainty relation, $\Delta x\Delta p> \hbar/2$.  For a mass $M$ with motion governed by non-relativistic force-free kinematics,  $\dot x= p/M$,
the standard quantum uncertainty of position difference measured at two times separated by an interval $\tau$ is\cite{caves1980a,caves1980,gardiner2004}
 \begin{equation}\label{quantum}
\Delta x_q(\tau)^2\equiv \langle (\hat x(t)-\hat x(t+\tau))^2\rangle> 2\hbar \tau/ M.
\end{equation}
This minimal uncertainty corresponds to a state with equal uncertainty from position and momentum.

Consider two bodies in  a quantum state of minimal  relative displacement uncertainty $\Delta x_q$.
The  uncertainty in their separation  is less than the change in  separation due to cosmic expansion in time $\tau$, $\Delta x_q< \tau H R$,
 if $M$ satisfies a lower bound,
\begin{equation}\label{quantumbound}
M>1/(\tau H) H R^2.
\end{equation}
The quantum uncertainty is minimized for a gentle measurement that takes place over a gravitational  time.
A determinate classical trajectory requires multiple samples in a single  orbit, so
$\tau H<1$, leading to the quantum  bound shown in Fig. (\ref{qc}). Larger masses are required to obtain approximately classical orbits.

\subsection{Quantum-Classical Boundary}

The lower bound on  size for an unperturbed cosmic system to behave classically is obtained where  gravitational and quantum bounds (Eqs. \ref{systemgravity} and \ref{quantumbound}) intersect (see Fig. \ref{qc}):
\begin{equation}\label{grainy}
R>H^{-3/5},
\end{equation}
a new scale of quantum indeterminacy associated with gravitational systems.  No measurement on a smaller scale can be made without disturbing the system.
On smaller scales, space is a quantum system with indeterminate geodesic trajectories; the motion and gravity of bodies are described by a spatially extended wavefunction with at least this width,  and motions of smaller subsystems are entangled with each other.

The corresponding system mass is  $M= H^{1/5}$.
For larger or smaller masses, or more than one measurement per orbit, the lower bound on size grows larger.
For the current mean cosmic density,  the  boundary  scale  is macroscopic: $H_0^{-3/5} \approx  60$ meters. 

Again, this scale has been derived only from standard non-relativistic quantum mechanics and gravity, and depends only on $H$, which is in turn fixed by the mean density of matter in the system.
The  approximations used to derive this result assume  classical geometry, so they break down on smaller scales. Indeed we do not know the quantum degrees of freedom of gravitational systems. However the approximations are valid up to the quantum-classical boundary, where they apply independently to each directional component of motion.  The distribution and motion of cosmic matter can have a homogeneous  wave function, but actual measurements of orbits on smaller scales yield large amplitude inhomogeneous and anisotropic departures from uniform expansion. No measurement of a homogeneous classical expansion is possible on smaller scales.  

In practical terms, such a measurement cannot be done close to any other masses because of their gravitational influence, and there is no place near  our solar system where the gravitational environment is sufficiently quiet for long enough, in a large enough region, to actually do the experiment.  However, the thought experiment is interesting because it reveals that  spatial locality, a foundational concept of space-time,  becomes ill defined at low curvature.

\subsection{Linear Perturbations}

Similar considerations can be applied to estimate the  quantum indeterminacy of  a more general system: a spatially extended, small amplitude, anisotropic and inhomogeneous  perturbation of uniform expansion. Consider a region of size $ R$, in which the velocity is perturbed by $\delta v$ from uniform expansion  in some direction, with a dimensionless amplitude  $\delta\equiv\delta v/(HR)<<1$.   In that direction, the classical displacement
$\Delta x_c$ of matter from the comoving frame over an interval $\tau$ is related to $\delta$ by
 $\Delta x_c=  R (\tau H) \delta$.  The dynamical  mass $M\approx  R^3H^2 $ is just the total mass in a region of size $R$.   Standard quantum uncertainty  (Eq. \ref{quantum}) then leads to 
 \begin{equation}\label{linear}
R> (\Delta x_q/\Delta x_c)^{-2/5} (\tau H)^{-1/5}\delta^{-2/5}H^{-3/5},
\end{equation}
which gives the same bound on $R$ as Eq. (\ref{grainy}),
since all the factors before $H^{-3/5}$ exceed unity for a linear classical perturbation. The linear theory thus also predicts nonlinear quantum fluctuations in orbital dynamics on this scale. 
 This bound again  does not invoke perturbed gravity explicitly:   perturbation dynamics  enters only through the kinematic relation $\dot x = p/ M$, and gravity only through the  background expansion rate $H$. It confirms that at any epoch, cosmological geometry is indeterminate  on a scale much larger than the Planck length.
 
Linear theory also shows that at high levels of precision (that is, at small $\delta$), classical solutions are somewhat  indeterminate quantum systems  for any $R$, although indeterminacy in $\delta$ is of course very small for large $R$. The amplitude of a linear perturbation  is   indeterminate  if  $\Delta x_q> \Delta x_c$.    Solving  for $\delta$ with   $\tau H< 1$ yields an estimate of the  width of the amplitude wave function at the quantum-classical boundary ($\Delta x_q=\Delta x_c$),
  \begin{equation}\label{amplitude}
\langle \delta^2\rangle^{1/2} \approx H  (RH)^{-5/2}.
\end{equation}
  This indeterminacy has a  negligible effect on large scale perturbations at late times, but it shows in a simple way how  basic quantum kinematics qualitatively  accounts  for perturbations  in the early universe.  The non-relativistic estimate should be valid for the center of mass motion of even relativistic matter at low velocities. It  predicts amplitude uncertainty $\delta \approx  H$  when extrapolated to the horizon scale $RH\approx 1$, which  is
  about the same as the amplitude of  metric  perturbations in  standard semiclassical inflation models based on effective field theory\cite{Lidsey:1995np,Lyth:1998xn}. Kinematic perturbations  of expansion in each direction are independent,  so  tensor and scalar components should have comparable amplitude, as indeed appears to be the case  for primordial perturbations   in  
 the real universe\cite{Ade:2014xna}.  Both  may be interpreted in this simple  picture as frozen-in imprints of  the  wave function of  early fluctuations, with an amplitude ($\langle \delta^2\rangle^{1/2} \approx 10^{-5}$) that depends on the mean value of $H$.  
 
 The magnitude of the uncertainty suggests that nonlocal correlations from new geometrical degrees of freedom  on scales smaller than $1/H$, but still much larger than the Planck length,  may influence the early quantum evolution of fluctuations.  
    

\section{Scale of dark energy}

These bounds  appear to be closely connected with the emergence of spatial locality in  cosmic systems.
Curvature is indeterminate in regions smaller than Eq. (\ref{grainy}), where  cosmic motions  are dominated by quantum noise. Presumably this   scale derives from the quantum system that gives rise to  space, gravity and quantum fields, along with cosmic expansion and acceleration. 



Non-local quantum entanglement of fields with geometry could connect the apparently unique cosmic acceleration scale $\approx H_0$ today with other scales of physics.
For example, consider the hypothesis\cite{cohen1999,Hogan:2013tza} that in a combined system of geometry and fields, there is a  maximum  size $R_{max}$ for  states of field modes  of frequency $M$,
\begin{equation}\label{validlimit}
R_{max}<  M^{-2}.
\end{equation}
This bound  constrains spatially extended states of excited fields  to be less massive than black holes, and  eliminates the need for  a  fine-tuned cancellation of  field vacuum energy on cosmic scales.
If field states non-locally ``notice''  gravitational states in this way,
it implies a relationship between gravitational curvature and particle mass,  such that field states extend at least as far  as the 
 quantum localization uncertainty for a particle of  mass $M$ over a gravitational time $\tau = H^{-1}$:  the two bounds (Eqs. \ref{quantumbound} and \ref{validlimit})  are consistent only  if 
\begin{equation}\label{lowerboundH}
H>M^3.
\end{equation}
This relation recalls the gravitational atom discussed above, as well as  the well-known\cite{zeldovich67,Bjorken:2010qx}  coincidence of current cosmic expansion rate with the scale of the strong interactions, $ H_0^{1/3}\approx  0.1$ GeV $\approx \Lambda_{QCD}$.  
 It could be that a minimum value of  $H$,  corresponding to the effect of a classical dark energy density or cosmological constant, is thereby set by the QCD vacuum, $M\approx \Lambda_{QCD}$.  If so, 
the  timescale associated with cosmic acceleration also  naturally  coincides with the  lifetime of stars\cite{Hogan:1999wh}.


\section{conclusion}
The cosmic  quantum uncertainty scales of length and perturbation amplitude (Eqs. \ref{grainy}, \ref{linear} and \ref{amplitude}) 
define   a quantum-classical boundary for geometry on scales much larger than the Planck length. In some regimes, the indeterminacy of the classical  metric could dominate semiclassical effects in models of dark energy or inflationary fluctuations  based on spatially averaged properties of matter, such as an energy-momentum tensor or Lagrangian density. The estimates here suggest that the dynamics of such systems may instead be largely shaped by properties of still-uncharacterized, spatially delocalized collective quantum states of matter and geometry.

\acknowledgments

I am grateful  for the hospitality of the Aspen Center for Physics, which is supported by  National Science Foundation Grant No. PHY-1066293.  
This work was supported by the Department of Energy at Fermilab under Contract No. DE-AC02-07CH11359.


 \end{document}